\documentclass[aps,prb,preprint,superscriptaddress,showpacs,preprintnumbers,amsmath,amssymb,floatfix]{revtex4-1}
\usepackage{graphicx}
\begin{document}

\preprint{}

\title{Quantum limited measurements with lossy optical cavity enabled by dissipative optomechanical coupling}

\author{Alexander K. Tagantsev}
\affiliation{Ioffe Physical-Technical Institute, 26 Politekhnicheskaya, 194021 Saint Petersburg, Russia}
\affiliation{ Swiss Federal Institute of Technology (EPFL), CH-1015 Lausanne, Switzerland}

\author{Sergey A. Fedorov}
\affiliation{ Swiss Federal Institute of Technology (EPFL), CH-1015 Lausanne, Switzerland}

\begin{abstract}
We analyze a cavity optomechanical setup, in which position of an oscillator modulates optical loss.
We show that in such setup quantum limited position measurements can be performed if the external cavity coupling rate matches the optical loss rate, a condition known as ``critical coupling".
Additionally, under this condition the setup exhibits a number of potential benefits for practical operation including the complete absence of dynamical backaction, and hence optomechanical instability, and rejection of classical laser noise and thermal fluctuations of cavity frequency from the measurement record.
We propose two implementations of this scheme: one based on signal-recycled Michelson-type interferometer and the other on a tilted membrane inside Fabry-Perot cavity.
\end{abstract}

\date{\today}

\maketitle

\newpage

\maketitle
\section{Paper}
Interferometric measurements are among the most precise methods to detect position of a macroscopic object and force exerted on it. This type of measurements is of great practical importance---it is employed in gravitational wave detectors~\cite{ligo2015}, microscopic on-chip sensors of mass~\cite{liu_sub-pg_2013} and magnetic field~\cite{rugar_mechanical_1992}, and acceleration~\cite{krause_high-resolution_2012}.
The sensitivity of interferometric measurements in realistic settings can closely approach the fundamental limit imposed by Heisenberg uncertainty principle, which provides experimental access to the quantum regime of measurement and control of mechanical motion in cavity optomechanics.
Impressive experimental progress has taken place in the measurement-based quantum cavity optomechanics in the recent years. Ground state cooling of a mechanical oscillator~\cite{wilson_measurement-based_2015,rossi_measurement-based_2018}, generation of squeezed light~\cite{safavi-naeini_squeezed_2013,purdy_strong_2013}, quantum-enhanced force metrology~\cite{kampel_improving_2017,sudhir_quantum_2017},
preparation of non-classical mechanical states~\cite{hong_hanbury_2017}
were demonstrated.

Experimental advances in quantum optomechanics so far have been relying on dispersive coupling---the  modulation of cavity resonance frequency by mechanical displacement. This coupling occurs naturally in a variety of systems and gives rise to coherent dynamics which enters the quantum regime when the radiation pressure shot noise exceeds the oscillator thermal force noise. Another conceptual alternative, however, is dissipative coupling when mechanical motion modulates optical coupling or loss rate~\cite{Elste2009}.
In this area the majority of research so far has concentrated on the former case---the modulation of optical escape rate via the driving and detection port %\cite{Elste2009,li_reactive_2009}.
It has remained unclear if quantum-limited operation is possible at all in the case when mechanical resonator introduces optical scattering or absorption. Indeed, in dispersive optomechanics quantum-limited operation is only possible if the decay rate of optical cavity is dominated by coupling to the detection port, which is straightforwardly understood from the no-lost-information principle\cite{Clerk2010}. In this letter we demonstrate that, counter-intuitively, the situation is different if the additional cavity decay rate is modulated by the mechanical element, ---here quantum-limited operation is possible and the no-lost-information principle can be fulfilled in the presence of finite optical power loss.
Our result provides a better insight into the possibility of attaining the quantum regime in the settings of recent experiments which demonstrated dissipative coupling between mechanical motion and optical losses~\cite{meyer_monolayer_2016,zhang_suspended_2014,Sawadsky2015}.

\begin{figure}
\includegraphics [width=0.95\columnwidth] {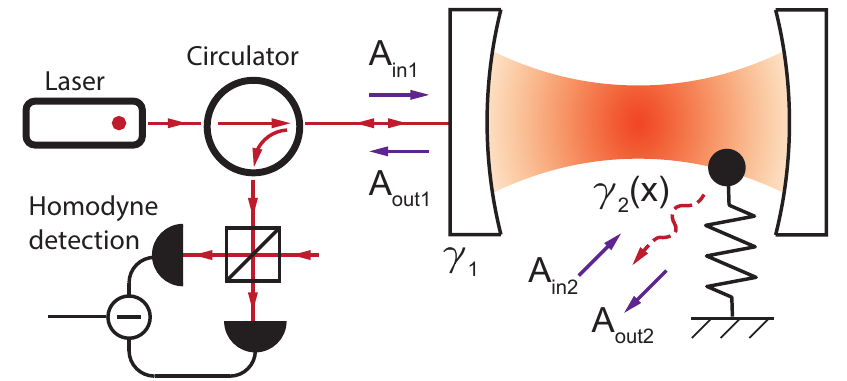}
\caption{Optomechanical setup under consideration.}
\label{fig:setup}
\end{figure}

We analyze an optomechanical setup consisted of (or equivalent to) a two-sided cavity in which one port is used for both excitation and detection and the other is not in use (see Fig.~\ref{fig:setup}).
The decay rate of the second port $\gamma_2$ is linearly coupled to the position of a mechanical oscillator.
Since optical losses are conventionally modeled by an undetected port, our scheme is equivalent to a cavity with mechanically modulated losses.
We show that the no-lost-information principle for position measurements is fulfilled here under the following conditions---operation in the unresolved sideband regime, ``critical coupling" ($\gamma_2$ equals the external coupling rate $\gamma_1$) and negligible dispersive coupling.
Additionally, we show that such a system never runs into optomechanical instability due to the absence of any dynamical backaction, provides laser noise rejection and yields signals not contaminated by the thermal cavity frequency noise.
In order to illustrate an application to quantum optomechanics, we show that the generation of perfect ponderomotive squeezing is possible despite the optical power loss due to the second cavity port.
The theoretical analysis is complimented by the discussion of realistic experimental implementations of our setup.

Optomechanical description of a one-sided cavity in which both the optical resonance frequency and the input coupling rate are linearly coupled to the position of a mechanical oscillator has been introduced by Elste et al~\cite{Elste2009} in terms of the quantum Langevin equations.
In this framework the incorporation of an additional optical port is straightforward\cite{Walls2008}. It leads to the following equations for the cavity field ladder operator $\textbf{a}$ and the operator of force $\textbf{F}$ acting on the mechanical oscillator

\begin{align}
\begin{aligned}
&\frac{\partial \textbf{a}}{\partial t}+\left[ \frac{\gamma_1+\gamma_2}{2}+i\omega_c\right] \textbf{a}
=\sqrt{\gamma_1}\textbf{A}_{\textrm{in1}}+\sqrt{\gamma_2}\textbf{A}_{\textrm{in2}}+
\\&\left[i g_{\omega0}\textbf{a}+  g_{\gamma0} (\textbf{a}-\textbf{A}_{\textrm{in2}}/\sqrt{\gamma_2})\right]\textbf{x}
\end{aligned}
\label{a}
\end{align}
\begin{equation}
\label{F}
\textbf{F}=\hbar g_{\omega0}\textbf{a}^\dag\textbf{a}+
 i\frac{\hbar g_{\gamma0}}{\sqrt{\gamma_2}} (\textbf{a}^\dag\textbf{A}_{\textrm{in2}}-\textbf{A}_{\textrm{in2}}^\dag\textbf{a}),
\end{equation}
where $\mathbf{x}$ is the oscillator position, $\hbar$  is Plank constant, $\omega_{c}$ is the cavity  resonance frequency, $\gamma_1$ and $\gamma_2$ are the cavity decay rates associated with the two ports, which are assumed to be much smaller than $\omega_{c}$. Also
 $ g_{\omega0} = -d\omega_{c}/dx$ and $ g_{\gamma0} =-(1/2)d\gamma_{2}/dx$ are the coupling constants of dispersive and  dissipative interactions, respectively.
$\textbf{A}_{\textrm{in1}}$ and $\textbf{A}_{\textrm{in2}}$
denote the photon flux-normalized optical input field operators for ports 1 and 2, respectively.

In our setup, the cavity is excited with a coherent laser drive via port 1 and the same port is used for detection, while the input of  port 2 is in vacuum state and its output is discarded.
We are interested in the fluctuations of $\mathbf{a}$ and $\mathbf{F}$ in the case when the dispersive coupling is absent ($ g_{\omega0} =0$).
In the reference frame rotating with the laser frequency $\omega_L$  the linearized frequency domain Langevin equations can be written as follows
\begin{equation}
\label{X}
\left[\frac{\gamma_1+\gamma_2}{2}-i\omega\right]\textbf{X}+\Delta\textbf{Y}
=\frac{\sqrt{\gamma_1}}{2}\textbf{X}_{\textrm{in1}}+
\frac{\sqrt{\gamma_2}}{2}\textbf{X}_{\textrm{in2}}+\mathfrak{G}_{\gamma}\textbf{x}
\end{equation}
\begin{equation}
\label{Y}
\left[\frac{\gamma_1+\gamma_2}{2}-i\omega\right]\textbf{Y}-\Delta\textbf{X}
=\frac{\sqrt{\gamma_1}}{2}\textbf{Y}_{\textrm{in1}}+\frac{\sqrt{\gamma_2}}{2}\textbf{Y}_{\textrm{in2}}
\end{equation}
\begin{equation}
\label{F1}
\textbf{F}=-\frac{\hbar\mathfrak{G}_{\gamma}}{\sqrt{\gamma_2}}\textbf{Y}_{\textrm{in2}}.
\end{equation}
where $\Delta=\omega_L-\omega_c$ and $\mathfrak{G}_{\gamma}= g_{\gamma0}a_0$, $a_0$ being the dimensionless amplitude of the driving field inside the cavity, set to be real.
We also used quadrature representation of the fluctuating parts of $\textbf{A}_{\textrm{in1}}$, $\textbf{A}_{\textrm{in2}}$, and $\mathbf{a}$.
Denoting their Fourier transforms as $\textbf{A}_{\textrm{in1}}(\omega)$ , $\textbf{A}_{\textrm{in2}}(\omega)$, and $\mathbf{a}(\omega)$, quadratures are defined as
$\textbf{X}(\omega)=[\textbf{a}(\omega)+\textbf{a}^\dag(-\omega)]/2$, $\textbf{Y}(\omega)=-i[\textbf{a}(\omega)-\textbf{a}^\dag(-\omega)]/2$,
$\textbf{X}_{\textrm{in1,2}}(\omega)=\textbf{A}_{\textrm{in1,2}}(\omega)+\textbf{A}_{\textrm{in1,2}}^\dag(-\omega)$,
$\textbf{Y}_{\textrm{in1,2}}(\omega)=-i[\textbf{A}_{\textrm{in1,2}}(\omega)-\textbf{A}_{\textrm{in1,2}}^\dag(-\omega)]$.
The correlators of the field quadratures satisfy the following relations
\begin{align}
\begin{aligned}
&\langle\textbf{X}_{\textrm{in1,2}}(\omega)\textbf{X}_{\textrm{in1,2}}(\omega')\rangle=
\langle\textbf{Y}_{\textrm{in1}}(\omega)\textbf{Y}_{\textrm{in1,2}}(\omega')\rangle
=
\\&i\langle\textbf{Y}_{\textrm{in1,2}}(\omega)\textbf{X}_{\textrm{in1,2}}(\omega')\rangle=
-i\langle\textbf{X}_{\textrm{in1,2}}(\omega)\textbf{Y}_{\textrm{in1,2}}(\omega')\rangle
\\&=\delta(\omega+\omega')
.
\end{aligned}
\label{XY1}
\end{align}
where $\langle ... \rangle$ stands for ensemble averaging.

The quadratures of the output fields leaving the cavity from port 1, $\textbf{X}_{\textrm{out}1}$ and $\textbf{Y}_{\textrm{out}1}$, can be found using the  standard  input-output relations~\cite{Walls2008}
\begin{equation}
\label{IOXO}
\textbf{X}_{\textrm{in}1}+\textbf{X}_{\textrm{out}1}=2\sqrt{\gamma_1}\textbf{X}\qquad \textbf{Y}_{\textrm{in}1}+\textbf{Y}_{\textrm{out}1}=2\sqrt{\gamma_1}\textbf{Y}.
\end{equation}
For port 2, the  input-output relation for the amplitude quadrature~\cite{Xuereb2011} should be modified to
\begin{equation}
\label{XY2}
\textbf{X}_{\textrm{in2}}+\textbf{X}_{\textrm{out2}}=2\sqrt{\gamma_2}\textbf{X}-\frac{2}{\sqrt{\gamma_2}} g_{\gamma0} \textbf{x}
\end{equation}
while that for the phase quadrature holds.

\textit{Wasted information.}
Keeping in mind the argument from Ref.\citenum{Clerk2010}, quantum-limited position measurements require the collection of all information imprinted by the oscillator on the optical field, as otherwise the measurement backaction from undetected signal increases noise above the minimum set by Heisenberg uncertainty principle.
In the case of our two-sided cavity quantum limited position detection can only be performed if the (undetected) output from port 2 carries no information about mechanical motion.
We show next that such situation can take place indeed.
At zero detuning ($\Delta=0$) it can be seen from Eqs.(\ref{X}), (\ref{Y}), and (\ref{XY2}) that the oscillator motion does not modulate the phase quadrature of outgoing light. In this  case,  the $\textbf{x}-$ dependent part of its amplitude $\textbf{X}_{\textrm{out2}}$ reads
$\textbf{X}_{\textrm{out2}}^{(x)} =2\gamma_2^{-1/2} \mathfrak{G}_{\gamma}\textbf{x}(\gamma_2 -\gamma_1 +2i\omega)/(\gamma_2 +\gamma_1 -2i\omega)$
revealing that, in the case of matching decay rates of two ports ($\gamma_2 =\gamma_1$) and bad cavity regime ($\omega\ll\gamma_1$), the output from port 2 contains no information about mechanical motion. This is remarkable given that under the same conditions all the optical power exits the cavity from port 2, which shows that the flow of information does not follow the flow of energy.

Interestingly, the same conclusion follows from a simple classical argument. In the bad cavity limit, the amplitude transmittance of our two-sided cavity is given by \cite{Walls2008}
$T =  2\sqrt{\gamma_1\gamma_2}/(\gamma_1+\gamma_2)$.
The dissipative optomechanical coupling introduces a dependence of $\gamma_2$ on the position of the oscillator $x$ at fixed cavity length.
Then, the sensitivity of the transmitted signal to mechanical displacement is given by $\frac{dT}{dx} =  \frac{\sqrt{\gamma_1}}{2}\frac{\gamma_1-\gamma_2}{\gamma_1+\gamma_2}\frac{d\gamma_2}{dx}$,
which is zero at critical coupling in accordance with the result of quantum treatment.

It is well known that in the case of dispersively-coupled optomechanical system with two-sided cavity the outputs of both ports always contain mechanical signal. The fact that dissipative coupling gives a different result can be understood as follows.
The variation of $\gamma_2$ due to dissipative coupling modulates the light twice: when it is inside the cavity and when it leaves the cavity (the latter is due to modified input-output relations). Under the condition of critical coupling these two contributions cancel each other.

\begin{figure}
\includegraphics
%[width=0.95\textwidth]{figures/plots.pdf}
[width=0.95\columnwidth]{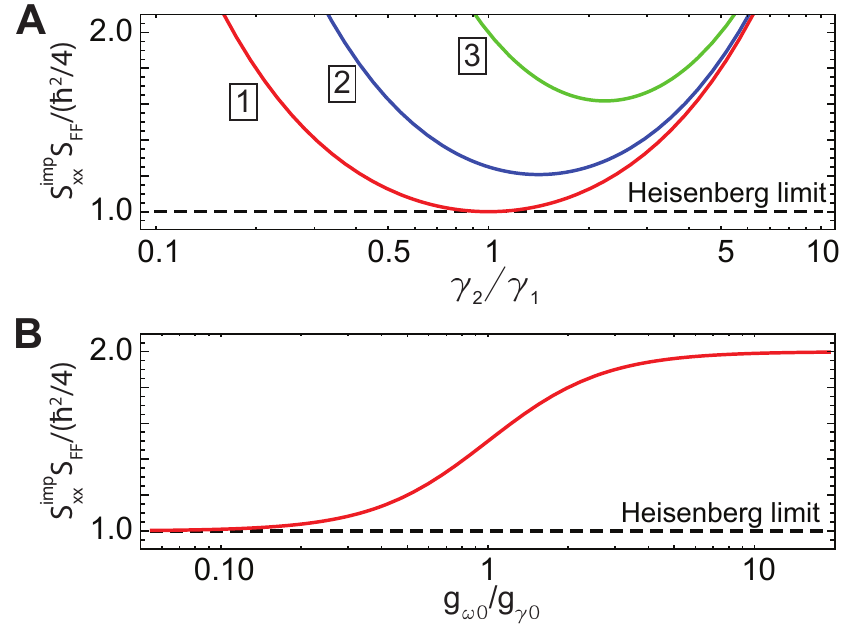}
\caption{Backaction-imprecision product for detection of output from port 1, A)  as a function of the second port coupling rate and the frequency in the rotating reference frame: 1 - $\omega/\gamma_1 = 0$, 2 - $0.5$, 3 - $1$; B) as a function of the dispersive optomechanical coupling constant
 $g_{\omega0}$.}.
\label{fig:plots}
%\end{figure*}
\end{figure}

\textit{Quantum limit in position measurements.}
Having found that under some conditions the output from port 2 contains no information about the oscillator, we next explicitly show that the uncertainty of position measurements using the signal from port 1 attains the quantum limit. It is the amplitude quadrature of the reflected signal $\mathbf{X}_{\textrm{out}1}$ that carries information about $\mathbf{x}$,  which reads
\begin{equation}
\label{OUT1}
\textbf{X}_{\textrm{out}1} =\frac{(\frac{\gamma_1-\gamma_2}{2}+i\omega)\textbf{X}_{\textrm{in1}}
+\sqrt{\gamma_1\gamma_2}\textbf{X}_{\textrm{in2}
}+
2\sqrt{\gamma_1}\mathfrak{G}_{\gamma}\textbf{x}
}{\frac{\gamma_1+\gamma_2}{2}-i\omega}.
\end{equation}
Using Eqs.(\ref{F1}) and (\ref{XY1}) we calculate~\cite{Clerk2010} the imprecision of position measurements as equivalent displacement noise power spectral density $S_{xx}^\textrm{imp}$ and the spectral density of quantum backaction force $S_{FF}$ as follows
\begin{equation}
\label{Sxx}
S_{xx}^\textrm{imp} =\frac{ \left( \frac{\gamma_1+\gamma_2}{2}\right )^2 +\omega^2}{4\gamma_1\mathfrak{G}_{\gamma}^2} \qquad S_{FF} =\frac{\hbar^2\mathfrak{G}_{\gamma}^2}{\gamma_2},
\end{equation}
implying the following product
\begin{equation}
\label{SQL}
S_{xx}^\textrm{imp} S_{FF} =\frac{\hbar^2 }{4} \frac{(\gamma_1+\gamma_2)^2 +4\omega^2}{4\gamma_1\gamma_2},
\end{equation}
which, at $\gamma_2 =\gamma_1$ and in the bad cavity limit ($\omega\ll\gamma_1$), attains the Heisenberg limit
$S_{xx}^\textrm{imp} S_{FF}=\hbar^2/4$.
This proves that quantum-limited measurements of position are possible by reading out only the signal from port 1.
Equation (\ref{SQL}) plotted in Fig.\ref{fig:plots}A yields the backaction-imprecision product as a function of the port decay ratio and sideband resolution factor.
Another important requirement for the quantum limit to be attainable, aside from the system being in unresolved-sideband regime, is the absence of dispersive coupling ($ g_{\omega0} =0$). The dependence of minimum backaction-imprecision product on dispersive to dissipative coupling ratio assuming $\gamma_1=\gamma_2\gg \omega$ is presented in Fig.\ref{fig:plots}B (see Supplementary Information for derivation). Here we observe the transition from one (no information is lost) to two (half of the information is lost) as $ g_{\omega0}$ is increased.
%
%\begin{figure*} \centering

%It is interesting to note that in case of one-sided cavity with either dispersive optomechannical interaction or dissipative interaction that modulates the input coupling rate (the situation considered to date in all available publications\cite{Elste2009,Xuereb2011,Weiss2013,Kilda2016,Tarabrin2013,Tagantsev2018}) the condition $S_{xx}^\textrm{imp} S_{FF} =\hbar^2/4$ is always satisfied, not only in the bad cavity limit in contrast to Eq.(\ref{SQL}).

\textit{Squeezing.}
Ability to perform quantum-limited position measurements is a prerequisite for a number of optomechanical experiments, including ground state cooling by feedback, quantum-enhanced metrology and the generation of squeezed light. As an example, we show that a two-port optomechanical cavity with dissipative coupling can generate squeezed light with the degree of squeezing limited only by the quantum backaction/thermal noise ratio.
Towards this end we introduce the dynamical equation for the mechanical oscillator driven by quantum backaction force $\mathbf{F}$ and thermal noise noise\cite{Elste2009}
\begin{equation}
\label{x2}
\chi^{-1}\frac{\textbf{x}}{x_{\textrm{zpf}}}=
\sqrt{\gamma_{\textrm{m}}}\textbf{W}+\frac{x_{\textrm{zpf}}}{\hbar}\mathbf{F}
\qquad x_{\textrm{zpf}}=\sqrt{\frac{\hbar}{2m\omega_{\textrm{m}}}}.
\end{equation}
Here  $m$, $\omega_m$, and $\gamma_m\ll\omega_m$ are the effective mass, resonance frequency, and damping of the oscillator respectively; mechanical force susceptibility is given by
$
\chi(\omega)^{-1}=\frac{1}{2\omega_{\textrm{m}}}\left[\omega_{\textrm{m}}^2-\omega^2 -i\gamma_{\textrm{m}}\omega\right].
$
The operator $\textbf{W}$ represents thermomechanical noise, its correlator is given by\cite{Habibi2016}
$
\langle\textbf{W}(\omega)\textbf{W}(\omega')\rangle\approx(n_{\textrm{\textrm{th}}}+1/2)\delta(\omega+\omega')$,
where $n_{\textrm{th}}$ is the thermal phonon occupation number of the oscillator (see Ref.\cite{Giovannetti2001phase} for the discussion of the approximation of thermal noise spectrum as white).

Using  Eqs.(\ref{X}), (\ref{Y}), (\ref{F1}), (\ref{x2}), and (\ref{IOXO}) one finds the quadratures of the reflected light as follows
\begin{align}
\begin{aligned}
&\textbf{Y}_{\textrm{out}1}=
\frac{\gamma_1-\gamma_2+2i\omega}{\gamma_1+\gamma_2-2i\omega}\textbf{Y}_{\textrm{in1}}
+\frac{2\sqrt{\gamma_1\gamma_2}}{\gamma_1+\gamma_2-2i\omega}\textbf{Y}_{\textrm{in2}}
\\&\textbf{X}_{\textrm{out}1}=
\frac{\gamma_1-\gamma_2+2i\omega}{\gamma_1+\gamma_2-2i\omega}\textbf{X}_{\textrm{in1}}
+\frac{2\sqrt{\gamma_1\gamma_2}}{\gamma_1+\gamma_2-2i\omega}\textbf{X}_{\textrm{in2}}+
\\&\frac{4\sqrt{\gamma_1\gamma_{\textrm{m}}}\chi(\omega)G_{\gamma}}{\gamma_1+\gamma_2-2i\omega}\left( \textbf{W}
-
\frac{G_{\gamma}}{\sqrt{\gamma_\textrm{m}\gamma_2}}\textbf{Y}_{\textrm{in2}}\right)
\end{aligned}
\label{reflection}
\end{align}
%
%$$\textbf{Y}_{\textrm{out}1}=\frac{\gamma_1-\gamma_2+2i\omega}{\gamma_1+\gamma_2-2i\omega}\textbf{Y}_{\textrm{in1}}
%+\frac{2\sqrt{\gamma_1\gamma_2}}{\gamma_1+\gamma_2-2i\omega}\textbf{Y}_{\textrm{in2}}$$
%$$\textbf{X}_{\textrm{out}1}=\frac{\gamma_1-\gamma_2+2i\omega}{\gamma_1+\gamma_2-2i\omega}\textbf{X}_{\textrm{in1}}
%+\frac{2\sqrt{\gamma_1\gamma_2}}{\gamma_1+\gamma_2-2i\omega}\textbf{X}_{\textrm{in2}}+$$
%
%\begin{equation}
%\label{reflection1}
%\frac{4\sqrt{\gamma_1\gamma_{\textrm{m}}}\chi(\omega)G_{\gamma}}{\gamma_1+\gamma_2-2i\omega}\left( \textbf{W}
%-
%\frac{G_{\gamma}}{\sqrt{\gamma_\textrm{m}\gamma_2}}\textbf{Y}_{\textrm{in2}}\right)
%\end{equation}
%
where $G_\gamma = g_{\gamma0} x_{\textrm{zpf}}a_0$.

We characterize squeezing by calculating the minimum of symmetrized (two-sided) power spectral density $S_{ZZ}$ of generalized quadrature
$
\textbf{Z}(\omega,\theta)=\textbf{X}_{\textrm{out1}}(\omega)\cos\theta+\textbf{Y}_{\textrm{out1}}(\omega)\sin\theta.
$
Calculations based on Eqs.~(\ref{reflection}) yield
\begin{equation}
\label{SZZfinal3}
S_{ZZ}(\omega,\theta)= 1 +{\cal L}(M\cos^2\theta + N\sin\theta\cos\theta)
\end{equation}
\begin{equation}
\label{M2}
M= 4n_{\textrm{\textrm{ba}}}\gamma_{\textrm{m}}^2|\chi(\omega)|^2(n_{\textrm{\textrm{th}}}+n_{\textrm{\textrm{ba}}} +1/2)
\end{equation}
\begin{equation}
\label{N2}
N= 4n_{\textrm{\textrm{ba}}}\gamma_{\textrm{m}}\textrm{Re}[\chi(\omega)]
\qquad{\cal L}=\frac{4\gamma_1 \gamma_2}{(\gamma_1+\gamma_2)^2+4\omega^2}
\end{equation}
\begin{equation}
\label{nba2}
n_{\textrm{ba}}= \frac{G_{\gamma}^2}{\gamma_{\textrm{m}}\gamma_2}.
\end{equation}
Here $n_{\textrm{ba}}$ is the oscillator excess phonon number occupation due to the quantum backaction of measurements, this quantity is also referred to as optomechanical cooperativity. Minimization over the quadrature angle at given frequency (see e.g. Ref.\cite{Tagantsev2018}) shows that, in a wide frequency range defined by the conditions
$
1 \ll|\omega_\textrm{m}-\omega|/\gamma_{\textrm{m}}\ll n_{\textrm{\textrm{th}}}+n_{\textrm{\textrm{ba}}},\, \omega_{\textrm{m}}/\gamma_{\textrm{m}},
$
the minimum of $S_{ZZ}(\omega,\theta)$ reads
\begin{equation}
\label{Sm}
S_{\textrm{m}}=S_{\textrm{0}}+(1-S_{\textrm{0}})(1-{\cal L}) \qquad S_{\textrm{0}}=\frac{n_{\textrm{\textrm{th}}} +\frac{1}{2}}{n_{\textrm{\textrm{ba}}}+n_{\textrm{\textrm{th}}} +\frac{1}{2}}.
\end{equation}
Equations (\ref{Sm}) predict that, at critical coupling ($\gamma_1=\gamma_2$), in the bad cavity limit ($\gamma_1\gg \omega_\textrm{m}$), and for large input power ($n_{\textrm{ba}}>>n_\mathrm{th}$), the minimum noise $S_\mathrm{m}\sim n_\mathrm{th}/n_{\textrm{ba}}$ goes to zero as $n_{\textrm{ba}}$ is increased.
This result is the same as for the generation of squeezed light in one-sided optomechanical cavity with either dispersive or dissipative coupling~\cite{Tagantsev2018}, the only difference being the definition of cooperativity $n_{\textrm{\textrm{ba}}}$.

\emph{Stability.}
A remarkable feature of optomechanical setup addressed is a complete absence of dynamical backaction and thus optomechanical instability.
This can be readily seen from Eq.(\ref{F1}) for the force acting on the oscillator, which does not have any contribution from the intracavity field, meaning that the \emph{dynamics of the oscillator is always decoupled from the cavity field}.
This is true for arbitrary detuning $\Delta$ and escape rate ratio $\gamma_1/\gamma_2$.

\emph{Rejection of laser and cavity frequency noises.}
It is interesting to note that the oscillator position measurements in the two-port configuration with dissipative coupling promise advantages over conventional dispersive readout in terms of classical noise rejection. First, according to Eqs.(\ref{X}) and (\ref{Y}) at zero laser detuning mechanical position fluctuations are imprinted on the amplitude quadrature of light field. This makes the signal free from the cavity frequency noises of thermal origin, including the mirror noise in Fabry-Perot cavities
\cite{braginsky_thermodynamical_1999,wilson_cavity_2012}
and thermorefractive noise in solid-state resonators
\cite{gorodetsky_fundamental_2004}.

Second, the operation at critical coupling ($\gamma_1=\gamma_2$) entails that, up to the sideband resolution factor $(\omega_m/\gamma)^2$, all classical noise of the driving laser (both phase and amplitude) exit the cavity from the second port and thus do not contaminate the detected signal. Also, different from the dispersively coupled optomechanical systems, here classical laser noise do not heat up the mechanical oscillator. This follows from Eq.(\ref{F1}), according to which only the input field from port 2, assumed to be vacuum, gives rise to the force driving the oscillator. These conclusions are remarkable, as they mean that the classical laser noise should have no impact on the position measurements in the setting that we consider. We make a reservation that this is only strictly true in the deep unresolved-sideband limit and assuming that the oscillator has no dispersive coupling in addition to dissipative.

\emph{Possible experimental implementation.}
The situation when dissipative optomecahnical coupling is larger than dispersive does not seem to be quite common. Nevertheless, strong dissipative coupling was already demonstrated in a few experiments. One approach is straightforward---placing a highly absorptive or scattering mechanical object in the optical cavity mode
\cite{zhang_suspended_2014,meyer_monolayer_2016}.
One problem with this approach is that the object may also introduce frequency shift and dispersive coupling, which is detrimental at least for the setup that we analyze in this letter. So far, in such systems, at best $ g_{\omega0}/ g_{\gamma0}=0.6$ was documented~\cite{meyer_monolayer_2016}.

In a different setup presented in
Ref.~\citenum{li_reactive_2009}
the flexural motion of a nanobeam modulates microdisk-waveguide coupling. Here the ratio of $ g_{\omega0}/ g_{\gamma0}=0.15$ was demonstrated, which is already quire promising and would in principle allow reaching the quantum limit in position measurements within just a few percent. Note that, in order to implement our setup shown in Fig.\ref{fig:setup},
one would need to add a second mechanically inactive waveguide that is coupled to the same microdisk, and use it for the laser drive. A disadvantage of integrated optical systems is the typically non-negligible intrinsic optical loss. As soon as this loss is not modulated by mechanical motion it does result in the loss of information about the mechanical oscillator.

Another interesting way to approach the problem of creating pure dissipative coupling is using a Michelson-Sagnac interferometer~\cite{Xuereb2011,Sawadsky2015,Khalili2016} with a reflective membrane inside. Here the phase modulation created by the membrane can be converted to effective reflectivity modulation via interference. While such dissipative coupling is of synthetic origin, it can be completely freed of the dispersive component by adjusting the membrane position\cite{Xuereb2011,Tarabrin2013}.

\begin{figure}
\includegraphics [width=0.95\columnwidth] {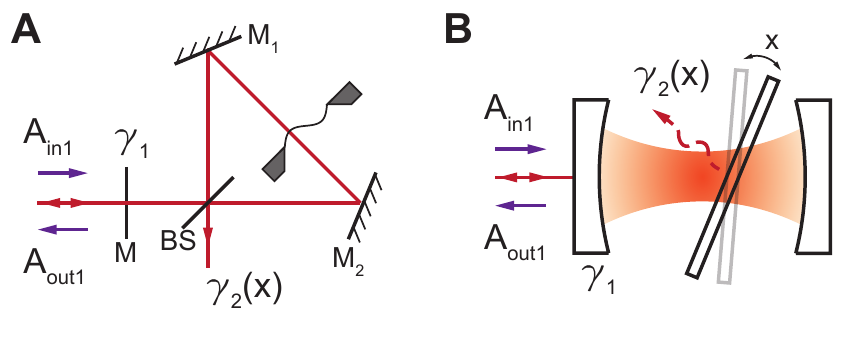}
\caption{Potential experimental realizations of two-sided optomechanical cavity dominated by dissipative coupling of undetected port.
A) signal-recycled Michelson-type interferometer. The light is coupled through the mirror M and back-reflected signal is detected. An effective second mirror is formed by the interferometer consisting of M$_1$, M$_2$ and BS, whose reflectivity is modulated by a position of the membrane.
B) One-sided Fabry-Perot cavity with tilted partially reflective element inside it that creates walk-off loss $\gamma_2$. Here the mechanical variable $x$ is the tilt angle. This system is dominated by dissipative coupling in the case of high-reflectivity of the middle mirror.
\label{devices}}
\end{figure}
Our two-port scheme with dissipatively-coupled loss port could be implemented by a minor modification of the apparatus used in~\cite{Sawadsky2015}, as shown in Fig.\ref{devices}A. Here the Sagnac interferometer formed by  mirrors M$_1$, M$_2$ and beamsplitter BS is seen from the outside as a single mirror with reflectivity proportional to the displacement of membrane. If another mirror (M) with reflectivity equal to that of the Sagnac part is added at the interferometer input, then the conceptual scheme presented in Fig.\ref{fig:setup} is perfectly reproduced. Interestingly, this two-port arrangement also enhances optomechanical cooperativity in unresolved-sideband regime compare to the setup used in\cite{Sawadsky2015}, which is equivalent to a cavity with a single dissipatively coupled port. According to Ref.\citenum{Tagantsev2018}, for a single-port cavity with dissipative optomechanical coupling cooperativity is given by
$
n_{\textrm{ba}1}= \frac{G_{\gamma}^2}{\gamma_{\textrm{m}}\gamma_2} \left( \frac{2\omega}{\gamma_2}\right )^2,
$
which much smaller than that of our two-port setup (given by Eq.(\ref{nba2})) by the sideband resolution factor $(2\omega/\gamma_2)^2$.

Finally, one more experimental platform that could implement our system is one-sided Fabry-Perot resonator with a slightly tilted semitransparent mirror inside (see Fig.\ref{devices}B). Here the middle mirror creates walk-off loss that can be also seen as coupling to higher-order optical modes of the cavity that have larger dissipation than the fundamental TEM$_{00}$ mode
\cite{sankey_strong_2010}.
Optical loss increases with increasing  the tilt of the middle mirror and in this way couples to mechanical motion if the mirror is free to rotate.
It should be noted that, in general, mirror rotations modulate the cavity resonance frequency together with dissipation,
however, as we show in Supplementary Information, the dissipative coupling dominates if the reflectivity of the middle mirror is high and it is placed half-way between the end mirrors.

To summarize, we analyzed an optomechanical cavity where the second non-detected cavity port is coupled to mechanical motion.
We showed that, counter-intuitively, the presence of dissipation here does not fundamentally prevent performing quantum-limited measurements of the oscillator position and the exploration of quantum optomechanics.
Such a lossy coupling may even bring advantages over the conventional dispersive coupling like the complete absence of optomechanical instability, measured signals in amplitude quadrature free of cavityfrequency noise and high rejection of the classical laser noises. Though the experimental implementation our model may be not straightforward, we discussed a few realistic configurations.

\section{Supplementary Material}
\subsection{Heizenberg limit controlled by combined  action of dissipative and dispersive coupling}
Let us  evaluate the backaction-imperfection product once both dissipative and dispersive coupling are active.
We will do it for the situation where $\gamma_1=\gamma_2=\gamma$ and $\omega/\gamma\Rightarrow0$.
In this case, the equation for the reflected  quadrature $\textbf{X}_{\textrm{out}1}$, Eq.(9) of the main text,
%Eq.(\ref{OUT1})
reads
\begin{equation}
\label{OUT1X}
\textbf{X}_{\textrm{out}1} =\textbf{X}_{\textrm{in}2} +a_0
\frac{2 g_{\gamma0}}{\sqrt{\gamma}}\textbf{x}
\end{equation}
while,  for the other quadrature $\textbf{Y}_{\textrm{out}1}$, staring from Eq.(1) of the main text, one readily finds
%(\ref{a})

%
\begin{equation}
\label{OUT1Y}
\textbf{Y}_{\textrm{out}1} =\textbf{Y}_{\textrm{in}2} +a_0
\frac{2 g_{\omega0}}{\sqrt{\gamma}}\textbf{x}.
\end{equation}
The optimal quantum-mechanical measurements must employ the  quadrature
$
\textbf{Z}_{\textrm{out}}=\textbf{X}_{\textrm{out1}}\cos\theta+\textbf{Y}_{\textrm{out1}}\sin\theta
$
such that the orthogonal quadrature carries no information about $\textbf{x}$.
Such a condition is met at $\theta =\tan^{-1}( g_{\omega0}/ g_{\gamma0})$.
For the optimal quadrature, we find
\begin{equation}
\label{OUT1Z}
\textbf{Z}_{\textrm{out}} =\textbf{Z}_{\textrm{in}} +a_0
\frac{2\sqrt{ g_{\gamma0}^2+ g_{\omega0}^2}}{\sqrt{\gamma}}\textbf{x}
\end{equation}
where the input noise $\textbf{Z}_{\textrm{in}}$ obeys  relations (6) of the main text
%(\ref{XY1})
implying the imprecision of position measurements
\begin{equation}
\label{SxxZ}
S_{xx}^\textrm{imp} =\frac{1}{4a_0^2}\frac{ \gamma}{ g_{\gamma0}^2+ g_{\omega0}^2}.
\end{equation}

In the situation considered,  a simple consideration started form Eq.(2) of the main text implies
\begin{equation}
\label{F1AD}
\textbf{F}=-a_0\frac{\hbar g_{\gamma0}}{\sqrt{\gamma}}\textbf{Y}_{\textrm{in2}}+2a_0\hbar g_{\omega0}\textbf{X}=
-\frac{\hbar a_0}{\sqrt{\gamma}}[ g_{\gamma0}\textbf{Y}_{\textrm{in2}}- g_{\omega0}(\textbf{X}_{\textrm{in1}}+\textbf{X}_{\textrm{in2}})],
\end{equation}
which leads to the following expression for the  spectral density of the quantum backaction force
\begin{equation}
\label{SFFZ}
S_{FF} =a_0^2\frac{\hbar^2( g_{\gamma0}^2+2 g_{\omega0}^2)}{\gamma}.
\end{equation}
Combining (\ref{SxxZ}) and (\ref{SFFZ}) we arrive on the following  backaction-imperfection product
\begin{equation}
\label{SQLZ}
S_{xx}^\textrm{imp} S_{FF} =\frac{\hbar^2 }{4} \frac{1+2\xi^2}{1+\xi^2}\qquad \xi=\frac{ g_{\omega0}}{ g_{\gamma0}}.
\end{equation}
Equation (\ref{SFFZ}) is plotted in Fig.2B of the main test suggesting that, for the configuration proposed,  the ability of reaching the Heisenberg limit is quite tolerant to moderate admixtures of the dispersive coupling.
\subsection{Coupling rates}
In order to calculate dispersive and dissipative coupling rates for the system shown in Fig.3B of the main text, we adopt a toy one-dimensional model that is physically relevant in an extreme case when high-order optical modes are so strongly damped that the light, once it has escaped from the fundamental mode, never returns back. Under this assumption the middle mirror can be described by scattering matrix with the sum of the power transmission and reflection coefficient smaller than one:
\begin{equation}
\label{LOSSmembrane}
 \left(
  \begin{array}{cc}
 it &  - r \\
- r  & it  \\
  \end{array}
\right),\qquad r=r_0+\delta r\qquad r_0^2 +t^2=1.
\end{equation}
Here $it$ and $r_0$ are the amplitude transmission and reflection coefficients of the middle mirror at normal incidence ($t>0$ and $r_0>0$) and $\delta r<0$ takes into account scattering to high-order optical cavity modes due to the tilt. By symmetry consideration $\delta r$ couples linearly to mirror rotation for a finite angle.
We adopt the scattering matrices
\begin{equation}
\label{mirrors}
\left(
  \begin{array}{cc}
 i\tau &  -\rho \\
-\rho  & i\tau \\
  \end{array}
\right)\qquad \textrm{and} \qquad
 \left(
  \begin{array}{cc}
 0 &  -1 \\
-1  & 0  \\
  \end{array}
\right),\qquad \rho^2 +\tau^2=1
\end{equation}
for the input mirror  and  end mirror of the cavity, respectively.
Using the matrices given by Eqs.(\ref{LOSSmembrane}) and (\ref{mirrors}), for the middle mirror put exactly half-way between the cavity mirrors, one readily finds the following equation for the resonant wavevector $k$ (complex in our case):
\begin{equation}
\label{condFULL}
(\rho^{-1}e^{-2ikl}-r)(e^{-2ikl}-r)+t^2=0
\end{equation}
In the absence dissipation, i.e. at $\rho=1$ and $r=r_0$, Eq.(\ref{condFULL}) implies the following well-known relation \cite{jayich2008}
\begin{equation}
\label{condFULL1}
\cos2k_cl=r_0
\end{equation}
for the real resonance wavevector of the system $k_c$.
In the general case, in term of the variable $X(r,\rho)$, solution to (\ref{condFULL}) reads
\begin{equation}
\label{X1}
X(r,\rho) =  \frac{r(1+\rho)\pm \sqrt{r^2(1-\rho)^2-4\rho t^2}}{2}
\end{equation}
where $\pm$ corresponds to two branches of the spectrum.
We are looking for the deviations $\delta k= k-k_c$ of the resonance wavevector form $k_c$, which are induced by nonzero values of $\delta\rho\equiv 1-\rho\approx\tau^2/2$ and $\delta r$ assuming $\delta\rho\ll1$ and  $ |\delta r|\ll1$.
For small $|\delta k|$, it can be expressed  in terms of $X(r,\rho)$ as follows
\begin{equation}
\label{deltak}
\delta k = - \frac{i}{2l}\left[1- \frac{ X(r,\rho)}{X(r_0,1)}\right].
\end{equation}
To assess the dissipation and resonance frequency shift induced by a small rotation of the middle mirror, we calculate the derivative of wavevector $\delta k$ with respect to $\delta r$ at $\rho=1$  to find
\begin{equation}
\label{correction1}
\delta k_r= \frac{\delta r}{2l}(\pm t\ +ir_0).
\end{equation}
Decomposing $\delta k$ given by Eq.(\ref{correction1}), we find the resonance  frequency shift
\begin{equation}
\label{frequency}
\delta\omega_c =c\textrm{Re}(\delta k) =\pm c\frac{\delta rt}{2l},
\end{equation}
and the decay rate
\begin{equation}
\label{decay}
\gamma_r =-2c\textrm{Im}(\delta k)= -c\frac{\delta rr_0}{l}.
\end{equation}
induced by the middle mirror rotation.

From Eqs. (\ref{frequency}) and (\ref{decay}) it follows that  the frequency modulation by tilt is smaller than the modulation of decay rate by a factor of $t/r_0$, and therefore dissipative coupling dominates the optomechanical interaction once the middle mirror is highly reflecting.

With $r=r_0$, Eq.(\ref{deltak}) describes the dissipation and variation  of the resonance frequency caused by finite $\delta\rho$ in the case of non-titled middle mirror.
Specifically, it readily reproduces the well-known expressions for the coupling rate of  the system
\begin{equation}
\label{decayrho}
\gamma_\rho =\frac{c\tau^2}{4l}
\end{equation}
and
\begin{equation}
\label{decayrho1}
\gamma_\rho =\frac{c\tau^2}{2l}.
\end{equation}
for the limiting cases of where $t/r_0\gg\tau^2$ and $t/r_0\ll\tau^2$, respectively.

Equation (\ref{deltak}) also enables us to set upper limit to the possible additional optomechanical coupling due to the interference between the effect of the mirror tilt and finite $\tau$.
It follows from this equation that, for the case of matching decay rates of the two ports where $|\delta r|\cong\tau^2$, such a coupling is smaller than dissipative one by a factor of $\tau^2$ and, thus, can be neglected.

We need make one remark more regarding the validity of our one-dimensional model.
The mirror rotations will, in general, change the transverse profile of optical mode due to the tilt-dependent hybridization of the fundamental and higher-order modes.
However, in the case when higher-order modes are highly dissipative, only a small admixture is required to create finite dissipation rate for the fundamental one, therefore we expect this effect to be also negligible within our assumptions.

\begin{acknowledgments}
The authors acknowledge fruitful comments  by Ivan V. Sokolov,  Eugene S. Polzik, and Farid Ya. Khalili.
The authors are also grateful to Eugene S. Polzik for reading the manuscript.
\end{acknowledgments}

\bibliography{QOwork,Serg}

\end{document}